**Title:** **The comparison of spectral analyses and flow features in upper airways with**

**obstructive sleep apnea (OSA) between successful and failed surgeries**

**Authors**: Mingzhen Lu[1]，Tianxi Chi[1], Zhengang Liu[2], Shu Wang[1], Yang Liu[1*] and

Jingying Ye[3*]

**Affiliation**

[1] Department of Mechanical Engineering, The Hong Kong Polytechnic University,
Kowloon, Hong Kong

[2] School of Power and Energy, Northwestern Polytechnical University, Xi'an, Shaanxi
Province, China

[3] Affiliated Tsinghua-Chang Gung Hospital, Tsinghua University, Beijing, China

**Contact information:**

Address for reprint requests and other correspondence: Y. Liu, Dept. of Mechanical

Engineering, The Hong Kong Polytechnic University, Hung Hom, Kowloon, Hong Kong

(Email: mmyliu@polyu.edu.hk)

**Abstract:**

Obstructive sleep apnea (OSA) is a common sleep disorder and widening the upper airway is often used in clinical practice. However, the success rate of this surgery is limited; the failed surgery would even make the situation worse, indicating the widened airway is not the unique criterion to evaluate the breathing quality. Therefore, we carried out both experimental measurement and numerical simulation on OSA upper airways and found that there existed an intrinsic dominant 3-5 Hz signal and the signal-to-noise ratio (SNR) at 3-5 Hz is inversely correlated with apnea-hypopnea index (AHI). Firstly, to validate the suitability of simulation methods, we carried out Laser Doppler measurement in 3D-printing OSA upper airway models, and found excellent agreement between the measured and calculated velocity profiles in two upper airway models for the first time. Then we carried out large eddy simulation (LES) to investigate four pairs of OSA upper airway models with 8 different AHI values for both pre- and post-surgery; among them, three surgeries were successful and one failed. The decreased the pressure drop for failed case, proving that only widening airway cannot guarantee to improve OSA. In our analysis, it is indicating that a dominant recirculation downstream of the minimum cross-section should be a main feature of a successful surgery, and the strength of 3-5 Hz signal induced by flow separation in the upper airway plays an important role in appraising breathing quality. This provides a new guideline for surgery planning. Furthermore, we found a strong correlation between AHI and the area ratio of minimum cross-section near the retro-palate to the maximum cross-section behind the tongue base, and this correlation is highly significant, $r = -0.833$, $p = 0.01 < 0.05$.

## 1. Introduction

Obstructive sleep apnea (OSA), which is usually caused by partial or complete narrowing of pharynx in the upper airway, is the most common type of sleep disorder (13). OSA are multifactorial and not just anatomically driven (29). There are various factors causing the collapse of airway, such as the abnormal anatomy of upper airway, pathological and insufficient reflex activation of upper airway dilator muscles and increased collapsibility of the passive upper airway. Among anatomical factors, airway narrowing has been reported wildly in OSA subjects. Therefore, fundamental to OSA treatment is delivery of positive airway pressure with salvage treatments directed towards advancing the jaw(s) or widening and stabilizing the airway. Based on the number of apneas and hypopneas per hour, the severity of OSA can be defined by the AHI (26). Uvulopalatophyngoplasty (UPPP) is the common surgical method to widen the airway; however the success rate of this surgery is limited. The post-operative complications could result in a dilemma during the operation of how much tissue to resect: too little would be ineffective, yet too much would make OSA worse (20). It seems the over-widened airway even makes AHI higher. Therefore, widened airway could not be the sole criterion to evaluate breathing, and it is necessary to study the other factors that would affect the breathing quality. It is known that oscillating pressure is the stimulus to trigger the reflex in the respiratory muscles (5). In our previous paper (7), we found that there exists an intrinsic dominant 3-5 Hz signal induced by flow separation in normal breathing for healthy subjects, we wonder whether this 3-5 Hz signal varies with AHI, if so, this 3-5 Hz signal may serve as a new criterion to evaluate the breathing quality.

Due to the complexity of upper airway, it is expensive and difficult to study the flow in upper airway experimentally, no matter in vivo or in vitro. Because of its non-invasive nature, the computational fluid dynamics (CFD) technique has been widely used to study the fluid flow in upper airway with OSA. In the early stage, only simplified upper airway models were used to conduct the CFD simulation (10) (11). Several studies were carried out in the realistic model of OSA upper airway based on CT/MRI images. Vos et al. (21) carried out CFD simulation using Reynolds-Averaged Navier-Stokes (RANS) method on 20 OSA upper airway models and studied the correlation between AHI and upper airway morphology. They reported that a combination of the smallest cross-sectional area and the resistance together with the body mass index (BMI) form a set of markers that predicted very well the severity of OSAHS in patients within their study. Nithiarasu et al. (15) carried out numerical simulation using RANS method based on CT-scanned upper airway model, and their numerical technique was validated against the measurement in an idealized human oropharynx model (4). Jeong et al. (6) studied numerically the flow in CT-scanned upper airway using low Reynolds number $k$–$\varepsilon$ model and found that the turbulent jet formed at the velopharynx due to area restriction was the most noteworthy feature in the pharyngeal airway of patients with OSA. Cheng et al. (2) also studied the flow in a realistic upper airway using an extended $k$–$\varepsilon$ turbulence model. However, the time averaged turbulence models cannot capture the flow features in the anisotropic flow such as adverse pressure gradients or turbulent velocity fluctuations in the upper airway (23). Hence, Large Eddy Simulation (LES) has been used in CFD simulation of flow in upper airway with OSA and has been validated as an accurate turbulence model by the experimental measurement (8) (14) (25) (28).

In this study, we carried out both measurement and LES simulation in upper airway models of four OSA subjects in which three of them had successful UPPP surgeries and one was failed. The calculated velocity profile on one model was validated against the laser Doppler anemometry (LDA) measurement. The objectives are to study the correlation between the 3-5 Hz signals induced by flow separation and the AHI, and to understand the effect of AHI on variation of flow patterns in OSA upper airway. These results may provide useful guidelines for surgical treatment planning.

## 2. Methodology

The investigation has been approved by the local ethics committee and is performed in accordance with the Declaration of Helsinki, and the subjects are provided with written informed consent forms.

### 2.1 Measurement on OSA subjects

Ten subjects were selected to carry out pressure measurement in the hypopharynx. The polysomnography study was carried out on these ten subjects as well to obtain the AHI, while the AHI of each subject is tabulated in Table 1. The measurement of pressure variation in the hypopharynx was conducted in Tsinghua-Chang Gung Hospital. The pressure catheter sensor (model MPC-500, Millar instrument Co, USA) was applied for detecting the upper airway pressure. There was a solid-state pressure-sensor at the tip of the catheter. The sensor was lying in the mouth/hypopharynx, superior to the vocal cord. The catheter was connected to the digital polygraphic system via a pressure transducer. The signals were observed and recorded by the additional channels of the PSG (polysomnography, Sandman Tyco, Ottawa, ON, Canada). The sampling frequency was 128Hz. The pressure sensor was calibrated at various pressures using a syringe attached to a manometer and by comparing the computer digital reading to the manometer reading within the range of 0 to 250 mm $H_2O$.

## 2.2 Reconstruction of 3-D upper airway model for simulations

Four severe OSA subjects were selected to carry out the simulation study, among them three had successful UPPP surgeries and one had failed surgery. The variation of AHI of each subject before and after surgery are tabulated in Table 2, where Subject #4 had failed surgery whose AHI increased from 46.1 to 81.7.

CT-scan was performed using a single-slice helical CT scanner (Phillips, Brilliance 64) in the affiliated Beijing Tongren Hospital, Capital Medical University. The images were taken when the subjects are awake. The two-dimensional images were obtained in the axial plane with a resolution of $0.7 \times 0.7$ mm$^2$, and slice thickness was 0.625 mm. We use the 2D region growing method (17) to segment CT images correctly. Then the three-dimensional point cloud data of upper airway models were reconstructed using the image processing software Mimics from nasal cavity to the laryngopharynx.

## 2.3 Experimental models

Two 1:1 scaled upper airway models, corresponding to subject #2 before and after UPPP surgery, respectively, were produced by 3-D printing technique (3500 HD Max-3D System, USA) with uniform wall thickness (0.5 mm). The printing material is VisiJet® M3 Crystal. The inlet and outlet were extended to minimize the velocity boundary effect. The LDA system (Dantec Dynamics, the uncertainty is about 2.5% at 0.15 m/s and 0.65% at 40 m/s) was utilized to measure the velocity profile inside the model. Since the laser beam diameter is 2.2 mm, the velocity within 3 mm of the wall cannot be captured. The seeding particles were smoke, generated by the Atomizer Aerosol Generator (TSI 3079) using DEHS oil. The air pump was connected with the airway outlet from tracheal side. The flow rate was controlled by a valve, and measured by flow meter (DryCal DC-2) ranging from 0 to 30 L/min, with an allowable deviation of 1%. Experiments were conducted at flow rate of 16.8L/min according to the bodyweight (8-10 ml/kg) (3).

A window is cut in the oropharynx where the wall is relative flat and smooth, it is covered tightly with the transparent thin film, and the window size is 15×15 mm for pre-surgery and 20×20 mm for post-surgery. Since the wall thickness is only 0.5mm, and transparent window would not affect the flow field significantly.

For the model before surgery, five cross sections with 1 mm spacing were chosen, and at each cross-section only one measuring line could be selected due to the narrowed airway. For the model after surgery, five cross-sections were chosen with 1 mm spacing, and at each cross-section three measuring lines were chosen with 1.5 mm spacing. The sampling time is 60 s for each point.

### 2.4 Numerical Methods

The flow field is solved by a CFD solver Fluent (ANSYS 14.5) with LES model. We assume the wall is solid. The air flow is assumed to be incompressible due to the very low Mach number (Mach < 0.3). The inlet and outlet were extended to minimize the velocity boundary effect, the inlet velocity profile follows the inspiratory curve with 12 cycles per minute, and the outlet pressure boundary condition is set to be zero. No-slip boundary condition is applied on the wall of airway and the time step is 0.001s. Second-order finite-volume schemes were employed for discretizing the flow government equations on the computational domain. The time-integration was performed using second-order implicit discretization, the coupling between the pressure and the velocity field was implemented through the SIMPLE algorithm.

The local Reynolds number ($Re=UD_{eq}/v$) ranged from 900 to 3200 according to the equivalent diameter ($D_{eq}$) of the cross-sectional area, the flow velocity ($U$) computed from the bulk flow rate and the kinematic viscosity of the air ($v$). The Reynolds number represents that the flow is from laminar to turbulent. Therefore, the WALE sub-grid scale (SGS) model is employed in the LES modelling because of its better ability of predicting the transition from laminar to turbulent regimes (22).

The mesh was generated using ICEM (ANASYS 14.5). The meshes are hybrid hexahedral/tetrahedral elements and a refined mesh was employed near the wall. Mesh convergency was tested by use of different mesh sizes until convergent to a prescribed tolerance (~0.2%).

### 2.5 Wavelet analysis and SNR variation

We used wavelet and fast Fourier transformation (FFT) to analyse the time series of pressure in hypopharynx. A wavelet is a function with zero mean and that is localized in both time and frequency domain (19). This feature allows us to determine both the dominant modes of signal oscillation and how those modes vary in time. Wavelet transform of a signal yields a three-dimensional structure above the time-frequency plane. Usually, the wavelet amplitude and power spectrum can be defined as the absolute values of the wavelet transform and their squares, respectively (19).

The power spectral density (PSD) exhibits the power at the certain frequency but it cannot indicate the quality of the signal. The signal quality can be appraised by signal-to-noise ratio (SNR), which is a measure to compare the level of a desired signal to the level of background noise. The SNR is defined in decibels (21) as

$$SNR = 20 log_{10} \left( \frac{A_s}{A_0} \right) \qquad (1)$$

where $A_s$ is the intensity of signal and $A_0$ is the intensity of noise background at the signal frequency.

## 3. Results

### 3.1 Inverse correlation between SNR at 3-5 Hz and AHI

Fig. 1 shows the wavelet analysis and SNR variation with AHI. From the wavelet analysis as shown in Fig. 1(a), at higher AHI or severe OSA, the respiration frequency exhibits beating pattern with time; at lower AHI or slight OSA, the breathing frequency is continuous with time. From the FFT analysis as shown in Fig. 1(b), besides the respiration frequency ($f_{re1}$), the second harmonic respiration frequency ($f_{re2}$) exhibits in the lower AHI subject, but disappears in the higher AHI one. Moreover, there exists an apparent signal ($f_{sep}$) at ~ 3-5 Hz range which is induced by flow separation near larynx. The SNR analyses show that the SNR decreases with increasing AHI at $f_{re1}$, $f_{re2}$, and $f_{sep}$ as indicated in Figs. 1(c) and 1(d).

### 3.2 The comparison between simulations and experiments

To validate the simulation results, we carried out the LDA measurement on the models of subject #2; for simulation, the inlet velocity profile follows the inspiratory curve with 12 cycles per minute as shown in Fig. 2(a). The velocity profile was measured along a straight line in a cross-section as indicated in Fig. 2(b). Figs. 2(c) and 2(d) show the comparison of axial velocities between calculated and measurement along the same straight line, and the calculated axial velocity contour indicates the location of data extracted.

For model before surgery (Fig. 2(c)), the velocity is higher near the posterior side due to the strong "pharyngeal jet" effect, and becomes weak and negative near anterior wall due to the induced reversed flow. As the laser beam diameter is 2.2 mm, LDA measurements cannot capture velocity in close proximity to the wall, but show the same profile and value as calculated over the entire measurable region. LDA measurement is particularly able to capture reversed flow near the anterior wall as indicated by the weak and negative velocity.

For the model after surgery (Fig. 2(d)), the calculated velocity profile is quite consistent with that of LDA measurement, and both captured region of separation and recirculation evident. From the comparison, we are confident that LES simulation is capable of capturing the flow characteristics in upper airway, particularly the reversed flow induced by separation.

### 3.3 The flow features in upper airways

In this section, all the discussions are based on the results of CFD simulations and the discussions are focused on the flow features in upper airways for subjects #1~#4 before and after surgery.

In upper airway with OSA, the most apparent feature for collapsing is a large negative pressure occurs in the minimum cross-section near the retro-palate (18). Fig. 3 shows that there exists a large negative pressure in the anterior side of the wall in the models before surgery for all subjects. The maximum pressure drop from choanae to the minimum cross-section is 490 Pa, 660Pa, 320 Pa and 340 Pa for subject #1 to #4, respectively, indicating a high flow resistance.

After the surgical treatment, as shown in Fig. 4., the large negative pressure is changed to positive near the retro-palate in the models of subject #1~#3 who had successful surgery. However, for the model of failed subject #4, the negative pressure still exists near the retro-palate and extended to the wall of oropharynx for both the anterior and posterior side. This would cause two collapse regions: one is near the retro-palate, and the other could be found in the oropharynx;the additional negative pressure in the oropharynx would make it worse. The maximum pressure drop from choanae to the cross-section near the retro-palate is 32 Pa, 50Pa, 23 Pa and 130 Pa respectively for subject #1 to #4. To quantify the results, the Spearman's correlation analysis was conducted, and a statistically significant correlation (1) was found ( $r = 0.738$, $p = 0.037 < 0.05$) between AHI and the maximum pressure drop from choanae to the cross-section near the retro-palate which consistent with the results of Wootton et al. (24) (Table 3).

Fig. 5 shows the axial velocity distribution at the sagittal-plane and cross-section at downstream of minimum area for OSA upper airway models before surgery. A strong jet-like axial velocity represented by dark blue colour develops behind the soft-palate for all subjects due to the anatomical narrowing of the upper airway. According to the Bernoulli equation, this high momentum axial velocity would result in a low pressure when the pressure reach to a critical value, it will change into a negative pressure as shown in Fig. 3. From contours at cross-sections, the high momentum flow occupies only small portion of the airway, and the induced reversed flow occupies most of the airway.

After surgical treatment, the jet flow in subjects #1~3 has been weakened significantly as a result of widening airway as shown in Fig. 6. From contours at cross-sections, the jet flow occupies most of the airway, and there is a single dominant reversed flow region which occupies less than half of the airway. However, the jet flow is still strong for subject #4 even it becomes weaker than that before surgery, and there are two main reversed flow regions due to the morphological change. For four upper airway models before surgery, the flow path-lines show that the high momentum jet flow induces several small and irregular recirculations.

Fig. 7 shows the wavelet analysis of flow time series in upper airway for normal subject and OSA subjects with both pre- and post-surgery. For the normal upper airway (AHI=4.5), it is clearly shown that there exists a strong signal around 3-5 Hz. For OSA subjects, there still exist signals around 3-5 Hz, but these signals are weak before surgery and become stronger after successful surgery. For better viewing, we extracted data from wavelet and shown in Fig. 7b. For OSA subjects #1-#3 with successful surgery, the 3-5 Hz signals are strengthened after surgery; for subject #4 with failed surgery, the 3-5 Hz signal becomes much weaker. Similar to measurement, an inspection of Fig. 8 also shows a strong inverse correlation between SNR and AHI for calculated signals at 3-5 Hz.

The morphology of upper airway affects the flow recirculation pattern, and the recirculation is reflected by the 3-5 Hz oscillation signal. However, in practical application, it is difficult to modify the flow recirculation pattern by defining the morphology. The feasible measurement is the cross-sectional area along the upper airway. We tabulated the AHI, minimum cross-section area near the retro-palate, maximum cross-section area in the oropharynx, theirs area ratio (AR) and AR variation in Table 3. We found a strong correlation between AHI and the area ratio of minimum cross-section near the retro-palate to the maximum cross-section behind the tongue base, and this correlation is highly significant (1), $r = -0.833$, $p = 0.01 < 0.05$, as shown in Fig. 9. For subject #4, the minimum area has been widened about 3.7 times, but the ratio between minimum area near the retro-palate and maximum area in the oropharynx does not change, indicating the ratio could be one of the characteristics of outcome of OSA surgery. It is worthy to note that there were no correlations between the increase in minimum cross-section area and AHI change, which is different from the correlation analysis in Mandibular advancement (MMA) (27).

## 4. Discussion

The widening upper airway can even decrease the pressure drop for failed case, indicating that widening airway cannot guarantee to improve OSA. In our previous study (7), we found experimentally an intrinsic peak frequency (3~5 Hz) for the normal subject resulted from the flow separation downstream the minimum cross-sectional area. Both the experimental and numerical studies show that the SNR at 3-5 Hz signal inversely correlates with AHI severity. This indicates that the 3-5 Hz oscillation signal induced by flow separation may play an important role in breathing control, as the oscillating pressure is the stimulus to trigger the reflex in the respiratory muscles (5). For most of the OSA subjects, the upper airway is narrowed which make the flow resistance increase; the widening airway may decrease the flow resistance but cannot guarantee to enhance the 3-5 Hz oscillation signal. That may be the reason of failed surgery for OSA subject.

Many researchers studied the flow in upper airway of OSA patients with CFD simulation, however, no one compared the difference between successful and failed surgery models and reported the importance of 3-5 Hz oscillation signal (9) (14) (16) (12). For models before surgery, the flow pattern is irregular, there are several recirculations in the reverse flow region; for models after successful surgery, the flow pattern is regular, and a single dominant recirculation is induced at the downstream of the minimum cross-section, and the oscillation profile of the wall shear stress time series is regular and follows the breathing curve closely (not shown in the paper), indicating that the regular oscillation profile is induced by the dominant recirculation. For the model after failed surgery, there are two main recirculations due to the morphological change, the oscillation of wall shear stress time series is strong but does not follow the breathing curve (not shown). Form the comparison of the flow patterns, we can find that upper airway surgery has improved the regularity of flow pattern of the reversed flow, i.e., in the model with lower AHI, the reversed flow pattern is more regular and there exists a single dominant recirculation at oropharynx.

The flow resistance or pressure drop has a significant effect on OSA. Wootton et al. (24) analysed numerically 15 pairs of obese OSA children, they found that the flow resistance in the pharynx and pressure drop from choanae to a minimum cross- section significantly correlated to the AHI, airway minimum cross-sectional correlation to AHI was weaker and the airway wall minimum pressure was not significantly correlated to AHI. Their conclusion is consistent with our result in this study. Further more, we also found a strong correlation between AHI and the area ratio of minimum cross-section near the retro-palate and the maximum cross-section behind the tongue base, $r = -0.833$, $p = 0.01 < 0.05$. Different from the correlation analysis in MMA by Zhao et al. (28), for UPPP surgery, there is no correlation between increase in minimum cross-section area and AHI change.

**Acknowledgements**

The support given by The Hong Kong Polytechnic University to MZ LU and TX Chi through the scholarships is gratefully acknowledged.

**Compliance with Ethical Standards:**

This study was funded by The Hong Kong Polytechnic University through scholarship (RPYG) to MZ Lu.

**Conflict of Interest:** Authors declare that they have no conflict of interest.

**Ethical approval:** All procedures performed in studies involving human participants were in accordance with the ethical standards of the institutional and/or national research committee and with the 1964 Helsinki declaration and its later amendments or comparable ethical standards.

pharyngeal airway model: An investigation of obstructive sleep apnea. *Journal of Biomechanics* 41: 2279-2288, 2008.

**Appendix**

**Computational Fluid Dynamics(CFD)**, is a branch of fluid mechanics that uses numerical analysis and data structures to solve and analyse problems that involve fluid flows.

 **Flow separation**－The fluid flow becomes detached from the surface of the object, and instead takes the forms of eddies and vortices

**Oscillation signal**- means flow-induced oscillation.

**Spectral output:**

**FFT analysis:**

The measured time series are analyzed by Fast Fourier Transformation (FFT). For a real signal $f$(t), if we regard it as an ergodic process, its autocorrelation is defined by:

, $$R(\tau) = \lim_{T \to \infty} \frac{1}{T} \int_{-T/2}^{T/2} f(t) f(t-\tau) dt$$

and its correlation coefficient is defined by:

$$r(\tau) = \frac{\lim_{T \to \infty} \frac{1}{T} \int_{-T/2}^{T/2} \left[ \left( f(t) - \mu \right) \left( f(t-\tau) - \mu \right) \right] dt}{\sigma^2}$$

where $\mu$ and $\mu$ are the mean and variance of the signal $f(t)$ and given by,

$$\mu = \lim_{T \to \infty} \frac{1}{T} \int_{-T/2}^{T/2} f(t) dt \text{ , and } \sigma^2 = \lim_{T \to \infty} \frac{1}{T} \int_{-T/2}^{T/2} \left[ f(t) - \mu \right]^2 dt \text{ .}$$

The power spectral density (PSD) can be obtained by imposing Fourier transform (FT) on $R(\tau)$,

. $$S(\omega) = \int_{-\infty}^{\infty} R(\tau) e^{-i\omega\tau} d\tau$$

On the other hand, if we impose on $r(\tau)$, since

$$r(\tau) = \frac{1}{\sigma^2}\, R(\tau) - \frac{\mu^2}{\sigma^2}$$

we can get

$$s(\omega) = \int_{-\infty}^{\infty}\left[\frac{1}{\sigma^2}\, R(\tau) - \frac{\mu^2}{\sigma^2}\right]e^{-i\omega\tau}\, d\tau = \frac{1}{\sigma^2}\, S(\omega) - 2\pi\frac{\mu^2}{\sigma^2}\, \delta(\omega)$$

where $\delta(\omega)$ is Dirac function. Therefore, if $\omega \neq 0$, $s(\omega)$ and $s(\omega)$ have only difference of a coefficient and consequently could be considered as the same one for analysis though $s(0)$ and $S(0)$ are very different. Thus, in this study we also call $s(\omega)$ as PSD.

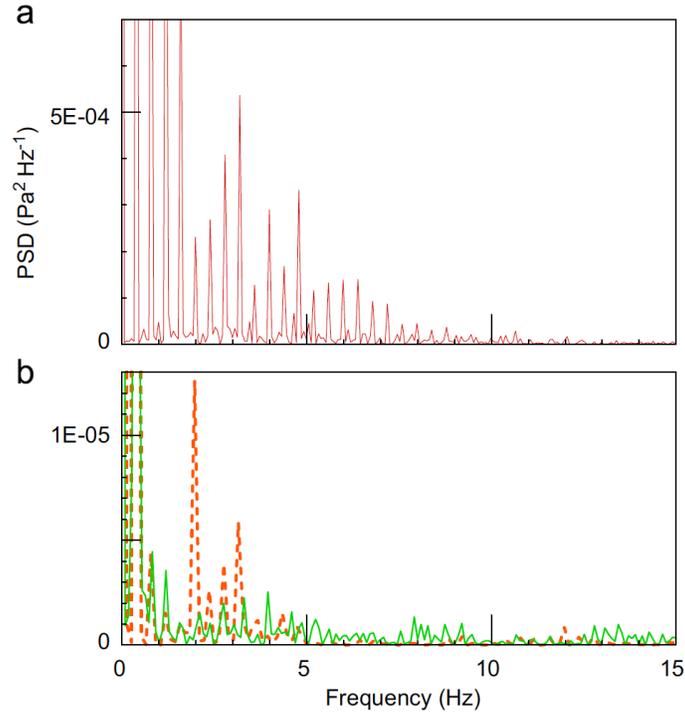

(a) Spectral analysis of wall shear stress time series at larynx for normal upper airway model. (b) Spectral analysis of wall shear stress time series at larynx for the #3 OSA model; solid line is for before surgery, and dashed line is for after surgery. It is shown that the PSD peaks around 3–5 Hz for these subjects. three models,.

**Wavelet transform:**

As in the windowed Fourier transform, one begins with a window function, which called a mother wavelet $\Psi(u)$. This function introduces a scale (its width) into the analysis. Commitment to any particular scale is avoided by using not only $\psi(u)$, but all possible scaling of $\psi(u)$. The mother wavelet is also translated along the signal to achieve time localization. Thus, a family of generally nonorthogonal basis functions:

$$\Psi_{s,t} = |s|^{-p} \psi(\frac{u-t}{s})$$

Where $p$ is an arbitrary nonnegative number. The prevailing choice in the literature is $p=1/2$. The use of scaled and translated version of a single function was proposed by Morlet in the analysis of seismic data. The continuous wavelet transform of a signal $g(u)$ is defined as:

$$\tilde{g}(s,t) = \int_{-\infty}^{\infty} \tilde{\Psi}_{s,t}(u) g(u) du$$

The wavelet transform $\tilde{g}(s,t)$ is a wavelet coefficient and $t$ is time, $s$ is the scale related to the frequency $f$ as $f = f_0/s$, and $f_0$ determines the current frequency resolution. By choosing $f_0 = 1$, we obtain the simple relation $f = 1/s$. The continuous wavelet transform is a mapping of the function $g(u)$ onto the time frequency plane. By adjusting the window used in wavelet transform, slower and faster events can be categorized accordingly. This method breaks down the steady fluctuating time series into its frequency elements and computes the power of signal components in predetermined frequency bands, allowing to measure the amplitude of different flow motion waves in PU/Hz.

The Apnea–Hypopnea Index or Apnoea–Hypopnoea Index (AHI) is an index used to indicate the severity of sleep apnea. It is represented by the number of apnea and hypopnea events per hour of sleep. The AHI values for adults are categorized as:

Normal: AHI<5

Mild sleep apnea: 5≤AHI<15

Moderate sleep apnea: 15≤AHI<30

Severe sleep apnea: AHI≥30

In this paper, we define successful surgery with the AHI decrease and the severe sleep apnea change into at least moderate sleep apnea after the UPPP surgery. However, for the failed surgery, the OSA subject still have severe sleep apnea and the AHI even increased.

**Table 1** AHI (episodes/hour) measurement of OSA subjects for the pressure experiment.

|  | AHI |  | AHI |
| --- | --- | --- | --- |
| Subject #1 | 3.5 | Subject #6 | 35.1 |
| Subject #2 | 5.2 | Subject #7 | 38.8 |
| Subject #3 | 8.5 | Subject #8 | 72.6 |
| Subject #4 | 19.7 | Subject #9 | 79.9 |
| Subject #5 | 23.7 | Subject #10 | 83.9 |

**Table 2** AHI (episodes/hour) measurement of OSA subjects for simulation.

|  | Before surgery | After surgery |
| --- | --- | --- |
| Subject #11 | 64.8 | 15.8 |
| Subject #12 | 60.7 | 23.9 |
| Subject #13 | 41.1 | 2.9 |
| Subject #14 | 46.1 | 81.7 |

**Table 3** AHI, minimum cross-section area near the retro-palate, maximum cross-section area in the oropharynx, theirs area ratio (AR) (min/max), the AR change (after/before) and the maximum pressure drop from choanae ($P_{choanae}$) to the cross-section near the retro-palate ($P_{min}$).

| | AHI | Min Area (mm$^2$) | Max Area (mm$^2$) | Area-Ratio (AR=Min/Max) | AR Change (After/Before) | $\Delta P$ (Pa) |
|---|---|---|---|---|---|---|
| Subject #1 | 64.8 | 53.2 | 256.4 | 0.21 | 2.00 | 490 |
| | 15.8 | 111.1 | 264.1 | 0.42 | | 32 |
| Subject #2 | 60.7 | 46.1 | 241.8 | 0.19 | 2.58 | 660 |
| | 23.9 | 318.1 | 645.0 | 0.49 | | 50 |
| Subject #3 | 41.1 | 66.7 | 149.2 | 0.45 | 1.67 | 320 |
| | 2.9 | 250.7 | 336.3 | 0.75 | | 23 |
| Subject #4 | 46.1 | 49.5 | 191.7 | 0.26 | 0.96 | 340 |
| | 81.7 | 92.6 | 370.9 | 0.25 | | 130 |

**Figure Captions**

**Fig. 1** The wavelet analysis and SNR variation with AHI: (a) the respiration frequency variation with time at two different AHI values; (b) the FFT analysis for three different AHI values;(c) and (d) the SNR analyses with AHI at the respiration frequency ($f_{re1}$), the second harmonic respiration frequency ($f_{re2}$) and the apparent signal ($f_{sep}$) at 3~5 Hz.

**Fig. 2** (a) The inlet velocity profile; (b) the sketch of the compared axial velocity line in a cross-section for subject #2; (c) and (d) the comparison between measured and calculated axial velocity profile for subject #2 before and after surgery respectively. The location on the measuring and simulating line is normalized as unity.

**Fig. 3** The wall pressure distribution in four models before surgery.

**Fig. 4** The wall pressure distribution in four models after surgery. The solid line in Subject #1 indicates the locations of pressure drop calculated.

**Fig. 5** The axial velocity distribution and path-line along the sagittal plane and the axial velocity contour of a cross section plane marked in the sagittal plane for subjects (a) #1, (b) #2, (c) #3 and (d) #4 before surgery.

**Fig. 6** The axial velocity distribution and path-line along the sagittal plane and the axial velocity contour of a cross section marked in the sagittal plane for subjects (a) #1, (b) #2, (c) #3 and (d) #4 after surgery.

**Fig. 7** The wavelet analysis of calculated flow time series in upper airway for normal subject and OSA subjects with both before and after surgery.

**Fig. 8** The correlation between AHI and SNR of calculated signal.

**Fig. 9** Correlation between AHI and the area ratio of minimum cross-section near the retro-palate and the maximum cross-section behind the tongue base ($r$ = - 0.868, $p$= 0.005 < 0.01).

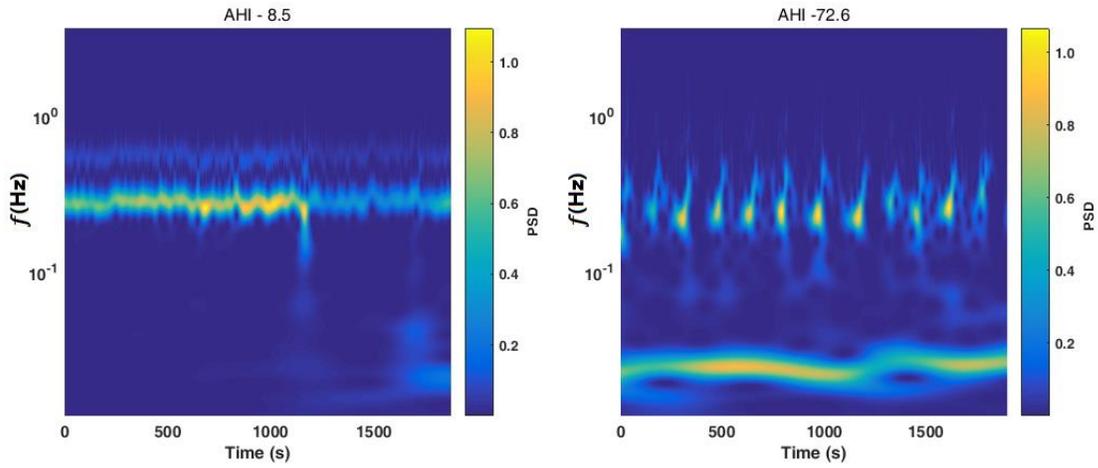

(a)

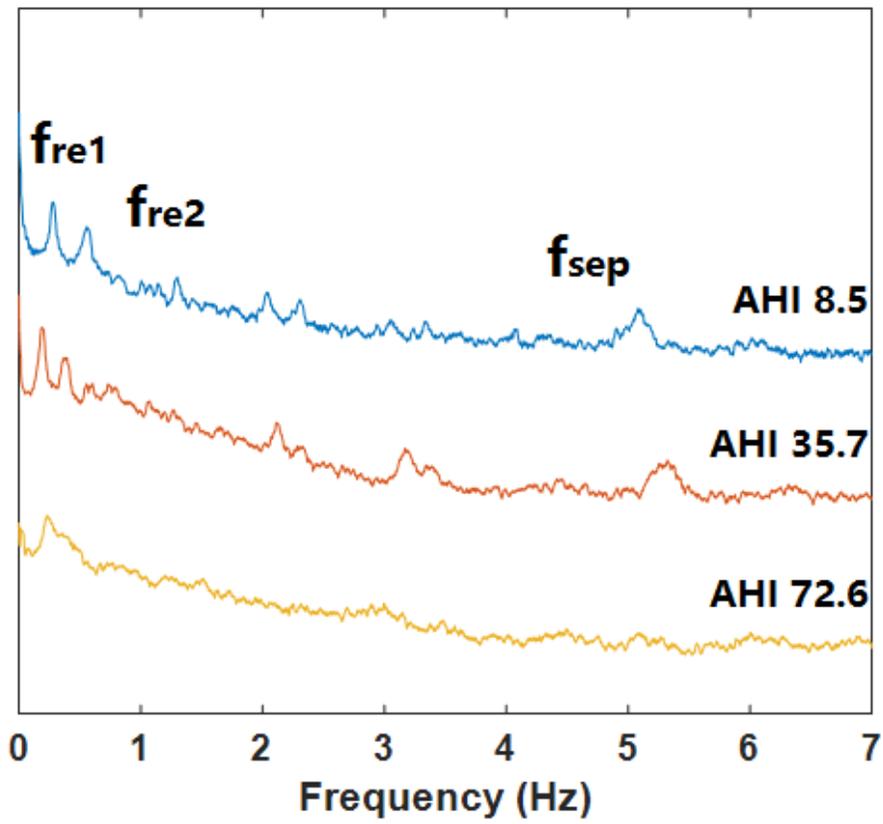

(b)

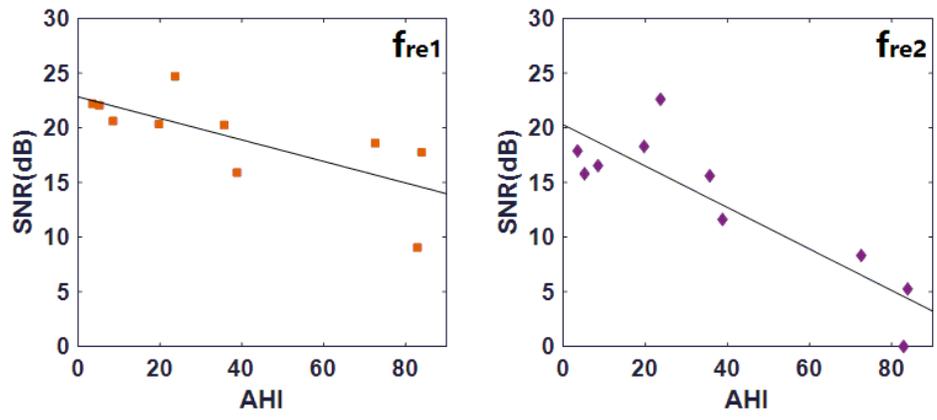

(c)

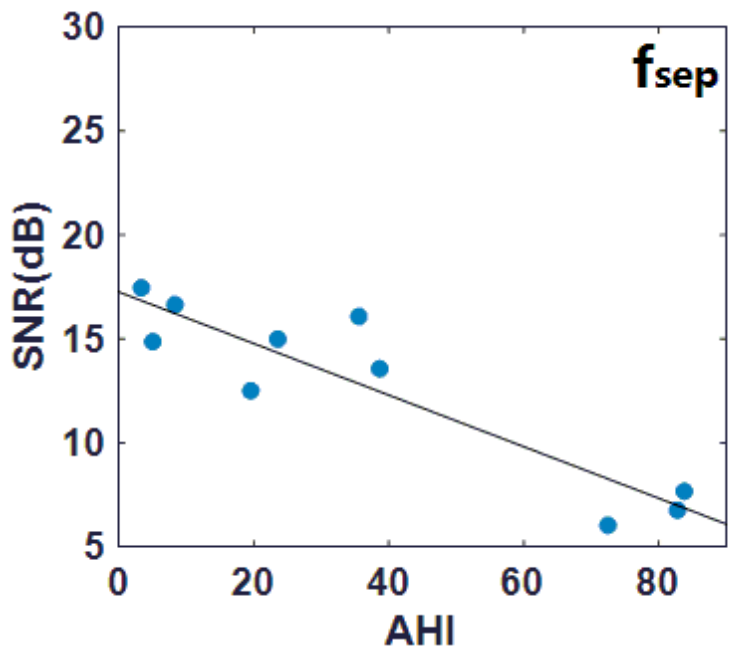

(d)

Fig. 1

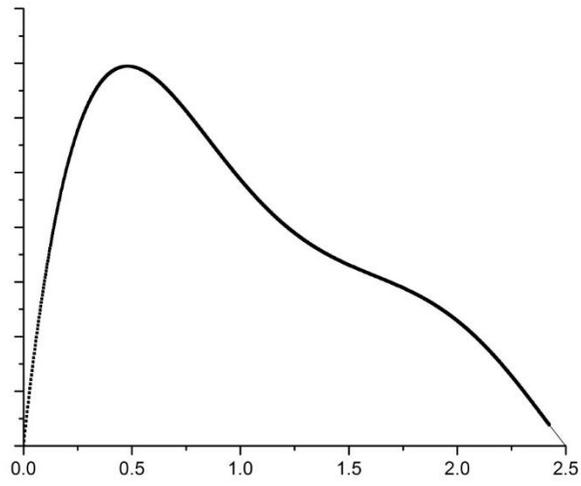

**(a)**

(b)

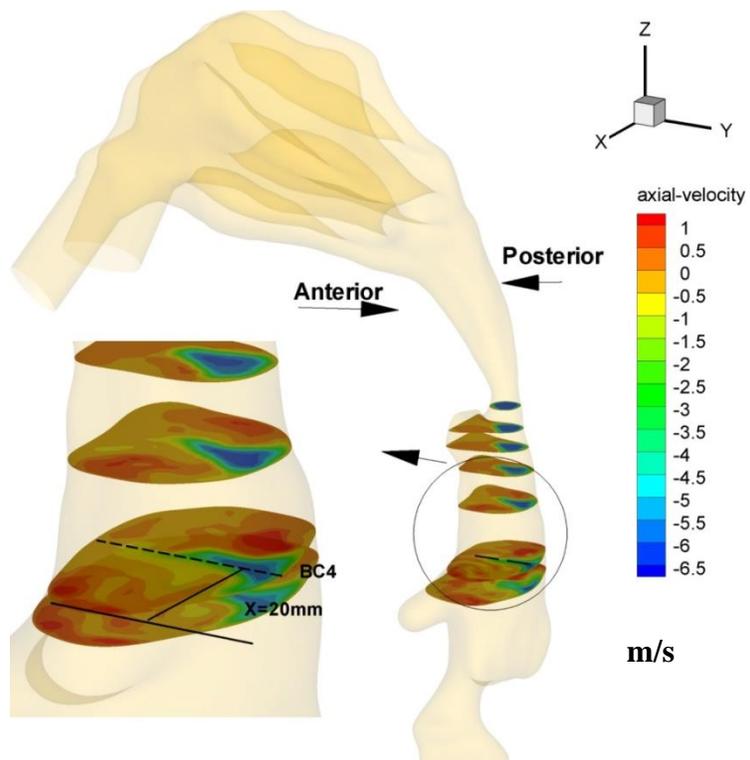

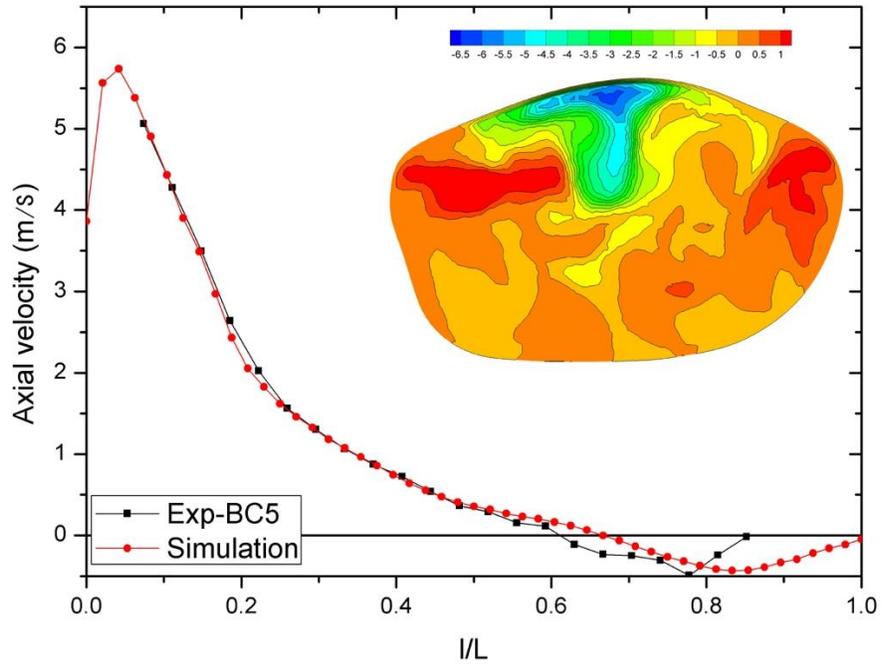

(c)

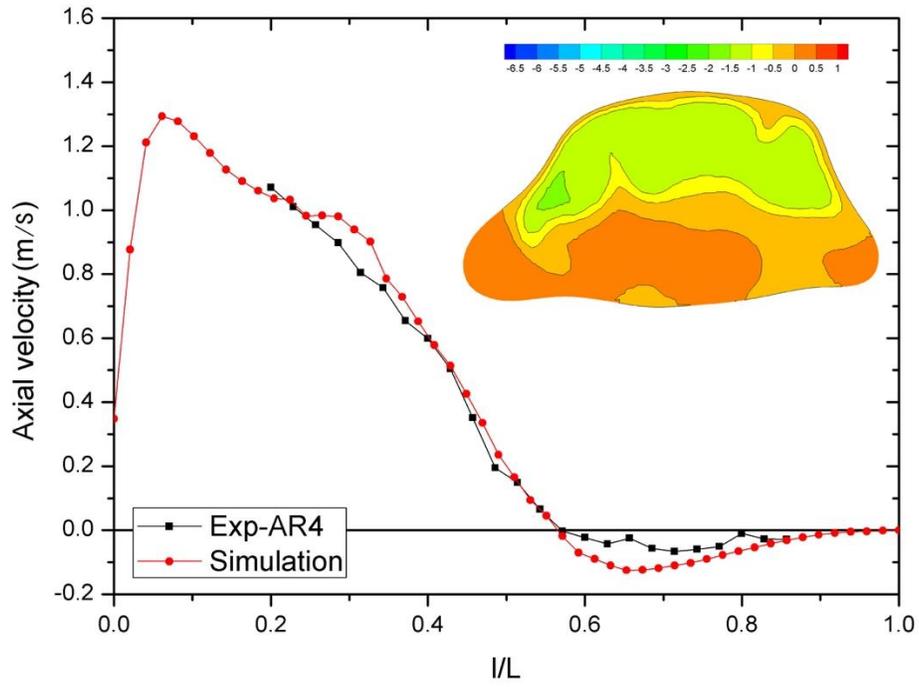

(d)

Fig. 2

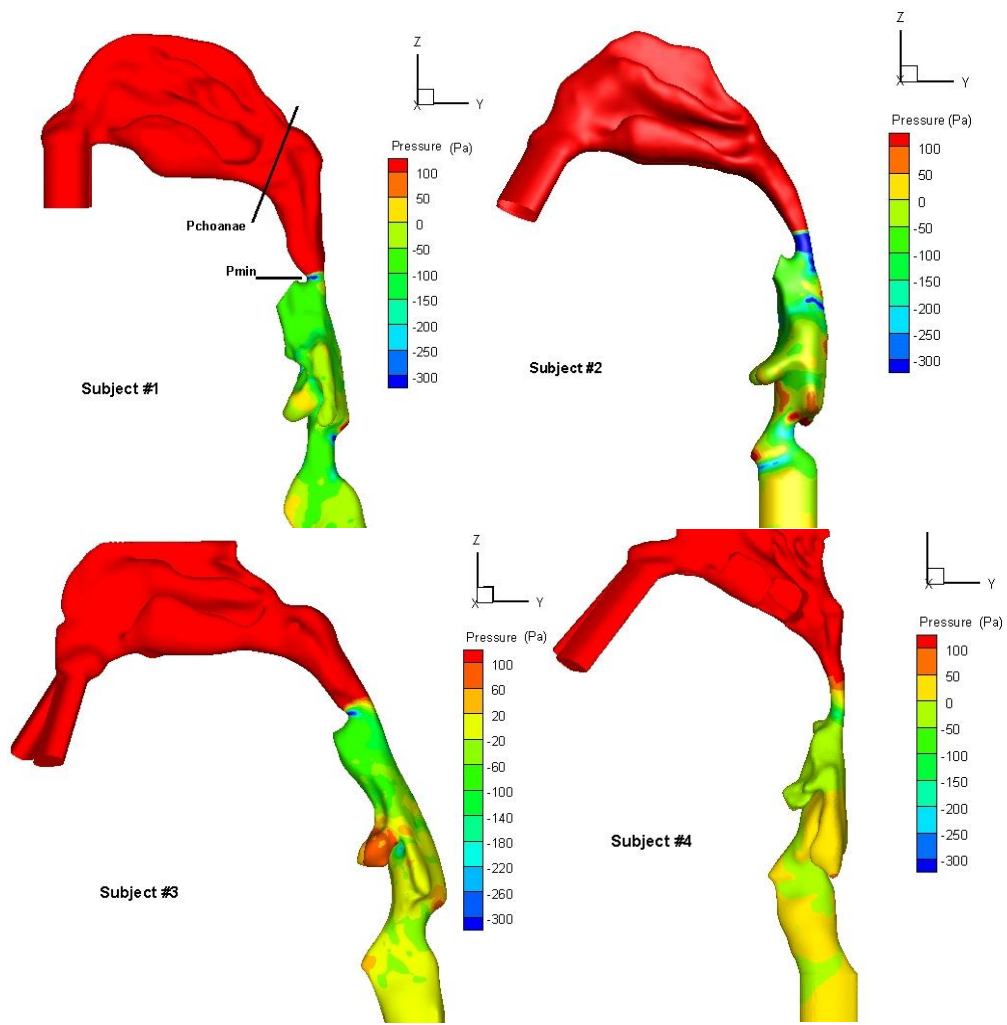

Fig. 3

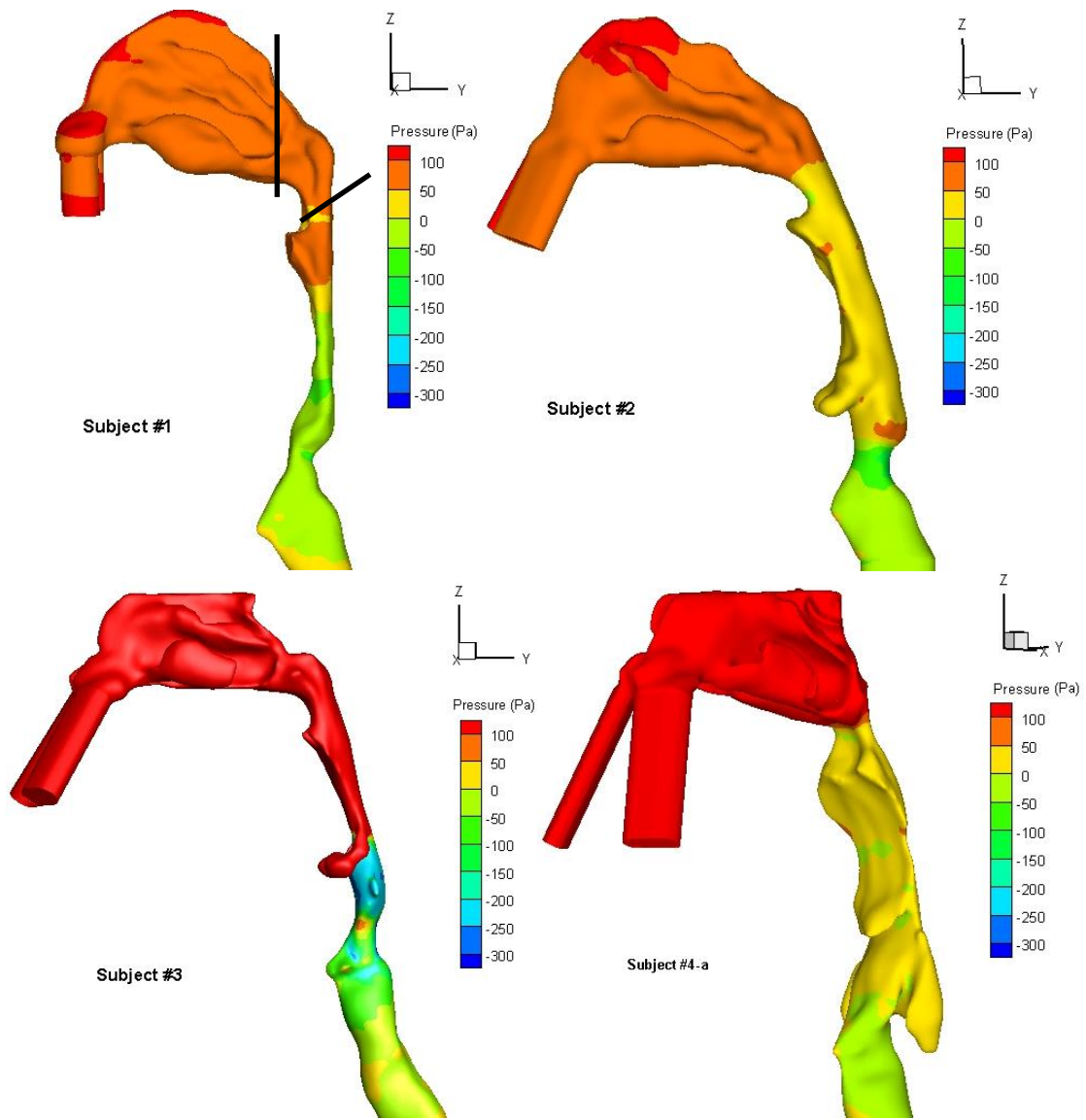

**Fig. 4**

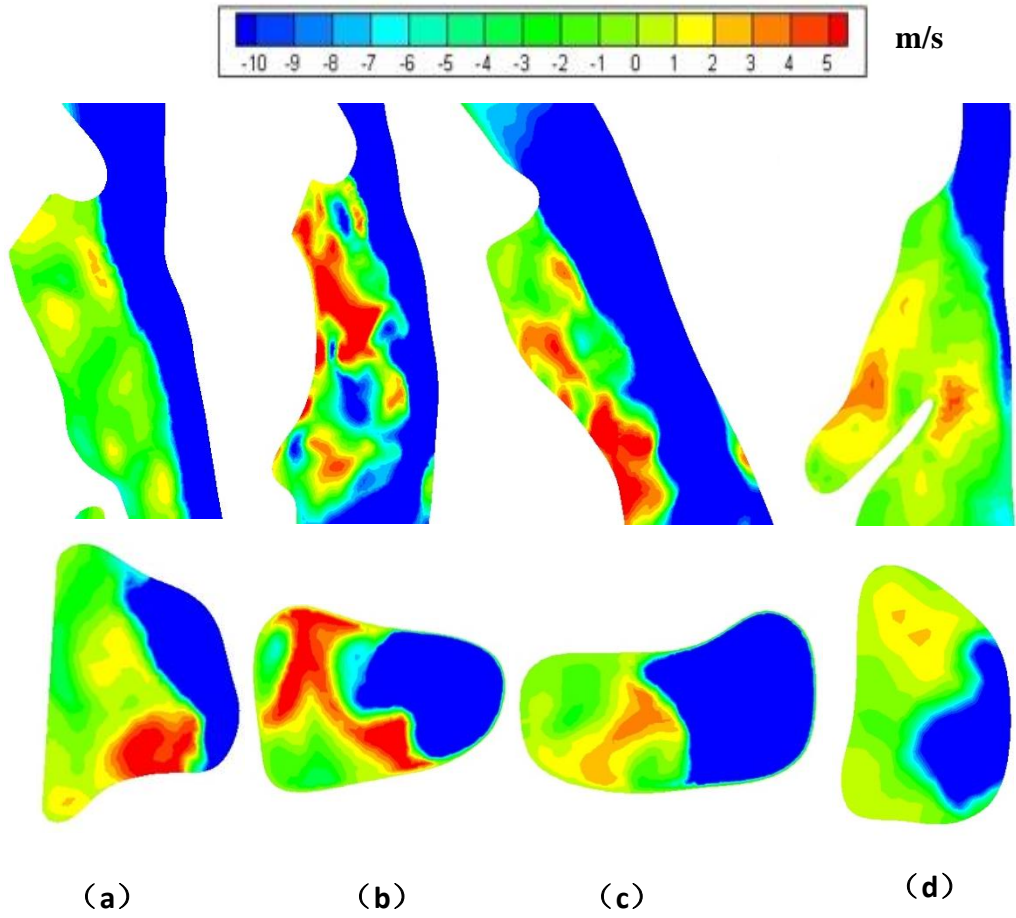

（a）        （b）        （c）        （d）

**Fig. 5**

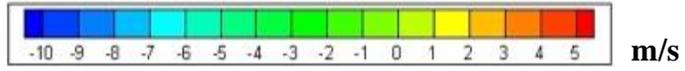

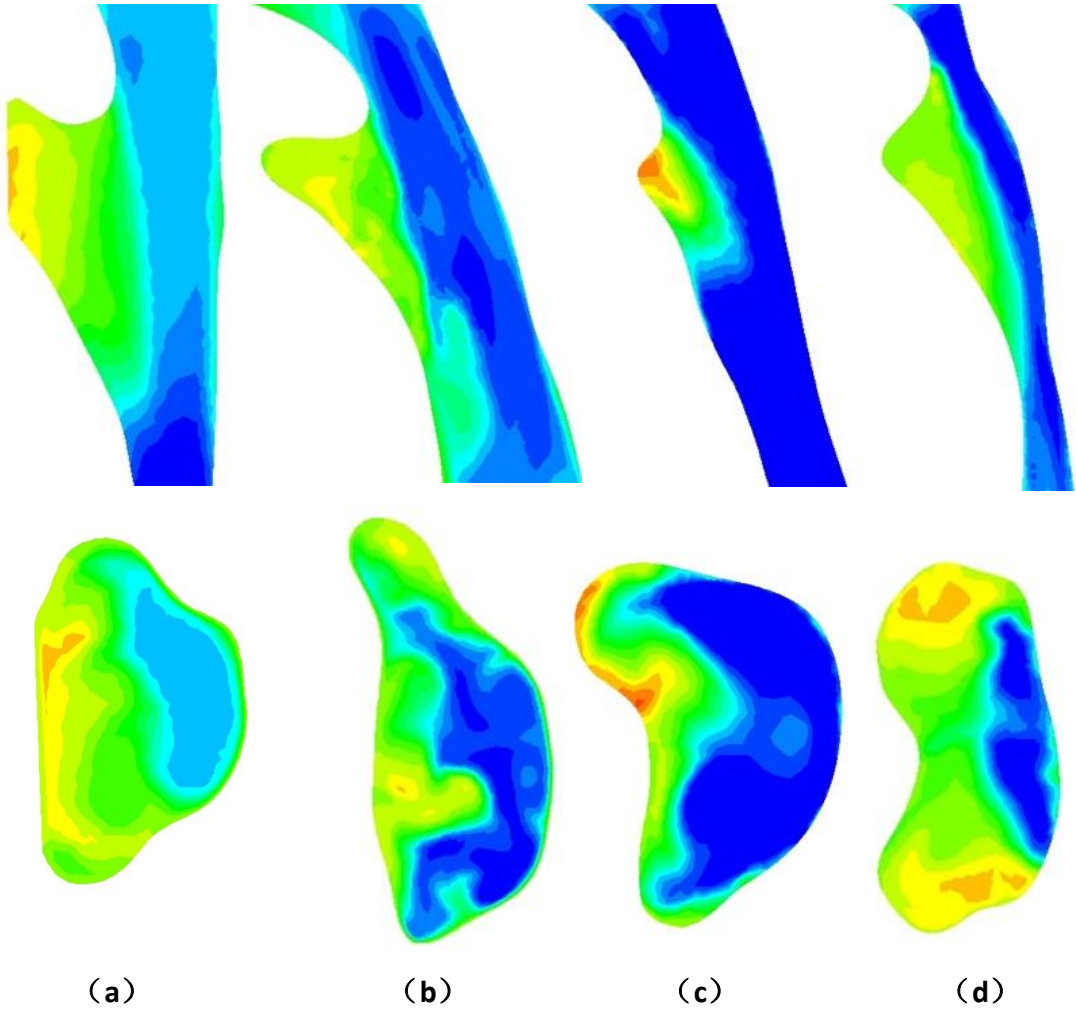

（a）        （b）        （c）        （d）

**Fig. 6**

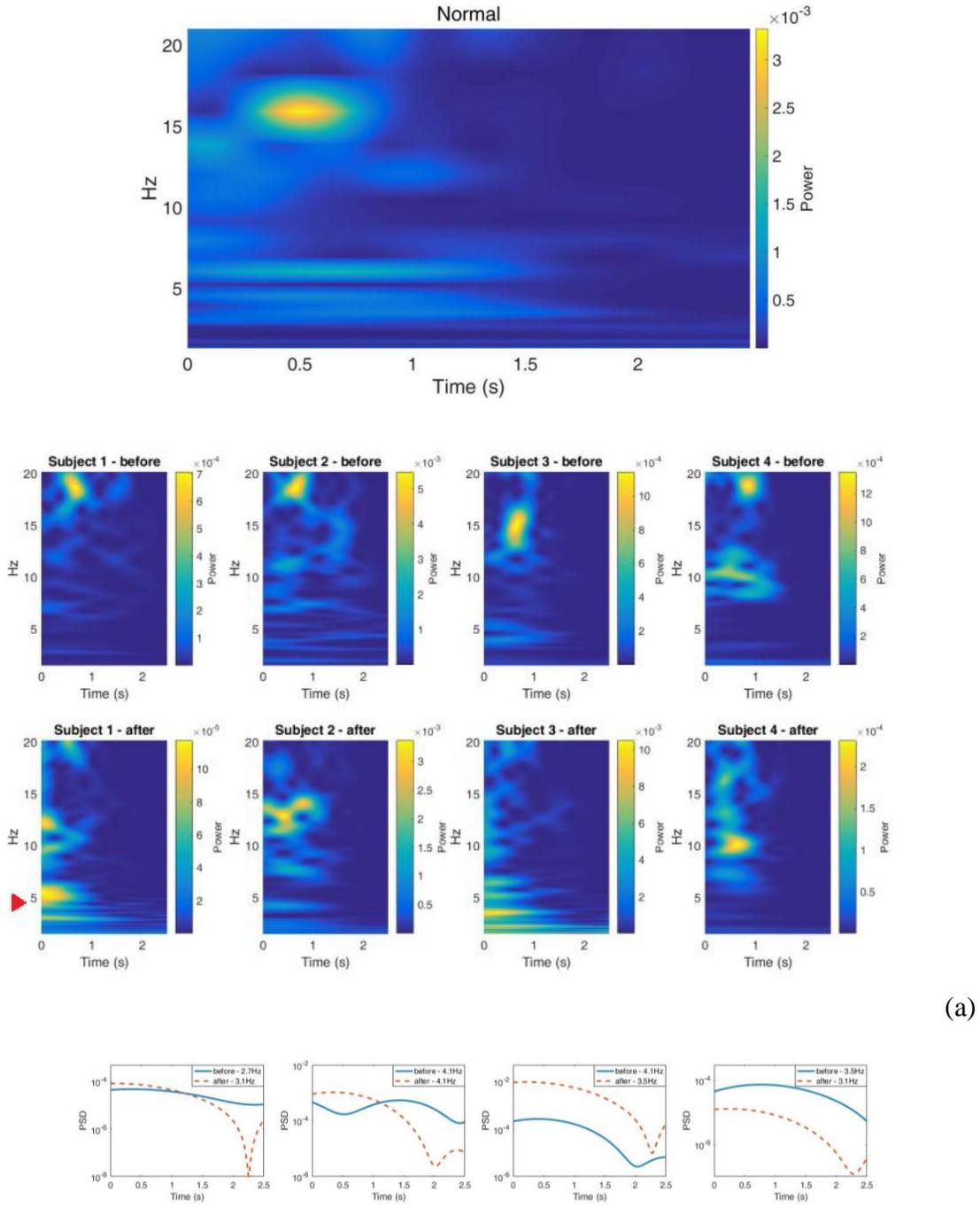

(a)

(b)

**Fig. 7**

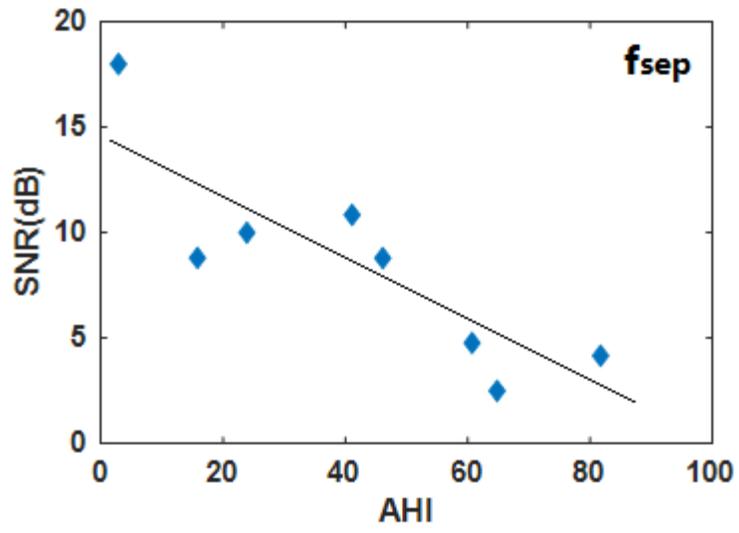

Fig. 8

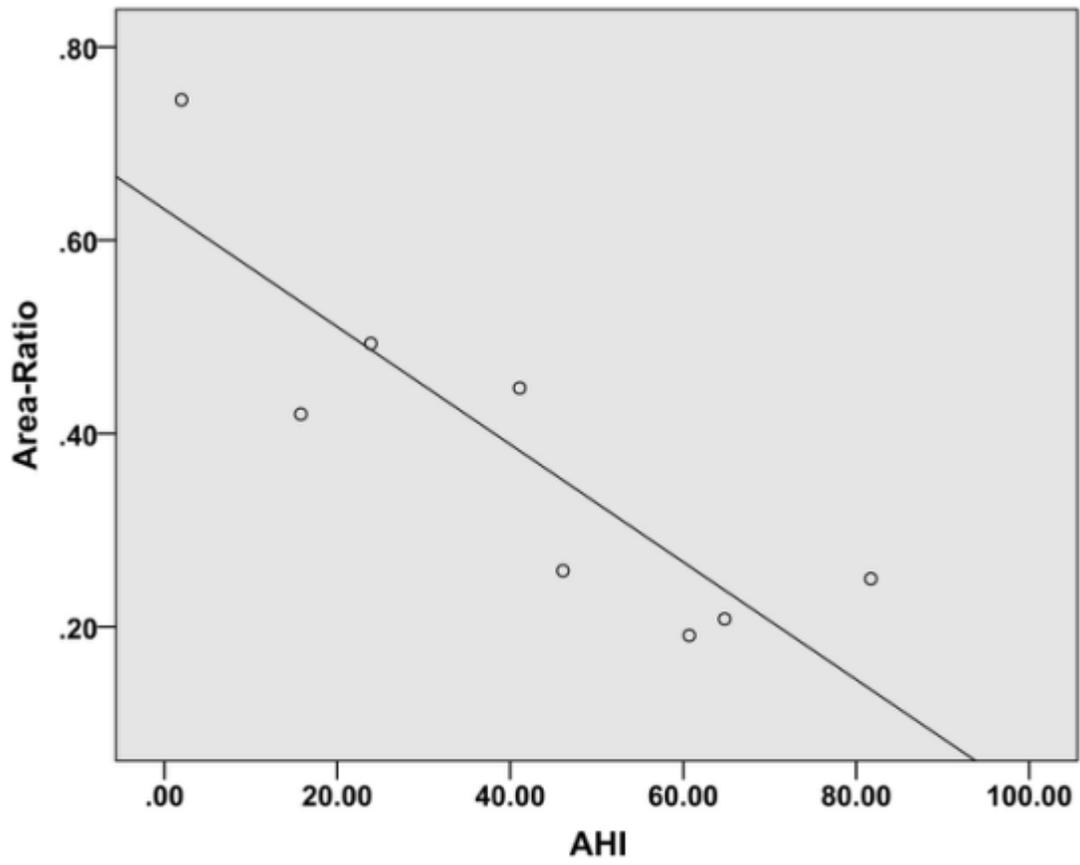

Fig. 9